\documentstyle [12pt,epsf]{article}

\input{epsf}
\begin{document}
\begin{titlepage}
\begin{center}

\vspace{-0.1in}

{\large \bf Single-Bubble Sonoluminescence as Dicke Superradiance\\ at Finite Temperature}\\
 \vspace{.3in}{\large\em M.~Aparicio Alcalde,\footnotemark[1], H. Quevedo\,\footnotemark[2]
 and N.~F.~Svaiter\,\footnotemark[3]}\\
\vspace{.2in}
Instituto de Ci\^encias Exatas e Tecnol\'ogicas, Universidade Federal de Vi\c{c}osa\,\footnotemark[1]\\
38810-000, Rio Parana\'{\i}ba, MG Brazil \\
\vspace{.2in}
Instituto de Ciencias Nucleares, Universidad Nacional Aut\'onoma de M\'exico\,\footnotemark[2]\\
AP 70543, M\'exico, DF 04510, Mexico\\
\vspace{.2in}
 Centro Brasileiro de Pesquisas F\'{\i}sicas\,\footnotemark[3]\\
 22290-180, Rio de Janeiro, RJ Brazil \\

\subsection*{\\Abstract}
\end{center}

\baselineskip .1in
Sonoluminescence is a process in which a  strong sound field is used to produce light in liquids.
We explain sonoluminescence as a phase transition from ordinary
fluorescence to a superradiant phase.
We consider a spin-boson model composed of a single bosonic mode
and an ensemble of $N$ identical two-level atoms. We assume that
the whole system is in thermal equilibrium with a reservoir at
temperature $\beta^{-1}$. We show that, in a ultrastrong-coupling regime, between the
two-level atoms and the electromagnetic field it is possible to have a cooperative interaction of the
molecules of the gas in the interior of the bubble with the field, generating sonoluminescence.

\vspace{0,1in}
PACS numbers: 05.30.Jp, 73.43.Nq, 78.60.Mq

\footnotetext[1]{e-mail: \,aparicio@ufv.br}
\footnotetext[2]{e-mail:\,\,quevedo@nucleares.unam.mx}
\footnotetext[3]{e-mail:\,\,nfuxsvai@cbpf.br}

\end{titlepage}
\newpage\baselineskip .18in

\section{Introduction}

\quad $\,\,$

Sonoluminescence is a process in which light is generated in a liquid under the action of a strong sound field.
 This phenomenon has been known since
the thirties of the last century \cite{sol1}. Gas bubbles trapped at the
pressure antinode of the sound field in a fluid are able to emit light  \cite{sol2, margulis}.
Single-bubble sonoluminescence was also reported in the literature \cite{sol3,sol4}.
The duration of the sonoluminescence pulse in single-bubble systems has been measured accurately.
Many authors stressed that the duration is
less then $50$ pc.
There are many phases in the process that can be summarized as follows. Due to the
strong sound field, cavitation occurs in the liquid.
The system absorbs energy and the radius of the bubble expands to a maximum value. The
time dependence of the bubble radius can be explained using the theory of cavitation or the
Rayleigh-Plesset equation \cite{sol5,sol6,sol7}.
The Raileigh-Plesset equation describing the dynamic of a pulsating bubble in an incompressible fluid can
be written as
\begin{equation}
r\,\frac{d^{2}r}{dt^{2}}+\frac{3}{2}\biggl(\frac{dr}{dt}\biggr)^{2}=-\frac{1}{\rho}\biggl
(p_{\infty}-p_{v}-p_{g}+\frac{2\sigma}{r}+\frac{4\mu}{r}\frac{dr}{dt}\biggr),
\end{equation}
where $r(t)$ is the bubble radius, $\rho$ is the density of the fluid, $\sigma$ and $\mu$ are
respectively the surface tension and the viscosity of the liquid.  The pressure $p_{\infty}$ corresponds to the liquid pressure at large distance from the bubble,
while $p_{g}$ and $p_{v}$ are the partial pressure of the gas and vapor in the cavitation bubble boundary, respectively. In the presence
of a sound field, the pressure of the liquid is given by
\begin{equation}
p_{\infty}=p_{h}-p_{m}\sin\omega t,
\end{equation}
where $p_{h}$ is the hydrostatic pressure and $p_{m}$ is the amplitude of the sound pressure field.
This picture is excellent in the explanation of the
expansion phase of the bubble as well as the preliminary stage of the collapse.
Since the bubble is immersed in a oscillatory pressure field, there is a subsequent compression phase.
During the collapse, the gas in the interior of the bubble is
compressed. The cavity collapse is supersonic, so a spherical convergent shockwave is driven \cite{sol8,sol9}.
The temperature of the gas in the interior of the bubble increases by the shockwave mechanism.
A region of high gas density and high temperature is created. Near the minimum 
a pulse of light is emitted. It is known that no sonoluminescence is produced unless the
gas in the interior of the bubble is doped with a noble gas. The intensity and temperature of
the sonoluminescence increases from $He$ to $Xe$.

There is no successful explanation for the mechanism involved in the conversion of phonons
of the liquid into ultraviolet photons. Some interesting open questions are related to the size of the
bubble when sonoluminescence occurs, the role of the inert gases in the process and
the mechanism of light emission. There are many different proposed mechanisms to explain the
light emission. In the shockwave model, in the last stage of the bubble collapse its
surface moves supersonically. The dynamic Casimir effect \cite{sol10,sol11} was used to explain the phenomenon,
but this approach was criticized and disregarded by many authors  \cite{sol12, sol13, sol14}.
Since the spectrum of the light is close to a thermal spectrum, brehmsstrahlung radiation was evoked
to explain the light emission \cite{sol15, sol16,sol17}. There are some points against this picture that we
would like to discuss. First, the bubble is a volume emitter, not a surface emitter in an
ideal blackbody. Also the emission of light from  the plasma cannot depend on whether the plasma was
generated from argon or other gas, but measurements show intensity differences.
An interesting point that we would like to stress is the fact that the wavelength of the
emitter light is very close to the minimum radius of the bubble. Other point is that the duration of light emission is quite short, as in 
the case of  Dicke superradiance.  We conclude that the mechanism of emission of light due to excited atoms and molecules
in the pulsating cavitation bubble can be explained as a cooperative phenomenon. An important question to answer is if the
emission of light in response to ultrasonic sound field is due to the same mechanism in single-bubble or many cavitation bubbles.

The aim of this paper
is to give a theoretical description of the phenomenon using Dicke superradiance
\cite{sol18,hioe,carmichael,duncan,sol19,yarunin, sol20, garraway}.
To this end, we consider $N$ identical two-level atoms prepared in the
excited state. For an ensemble of atoms in a
volume small compared to the emission wavelength, they start to
radiate spontaneously much faster and stronger than the ordinary
fluorescence case, where the radiation rate becomes
quadratic-dependent on the number $N$ of atoms. Other important
characteristic of this cooperative process is that the emission
has a well-defined direction depending upon the geometry of the
sample.

A summary of results obtained in the present literature is in order. In the Dicke
model, an important result was obtained by Hepp and Lieb
\cite{hepp}. These authors presented the free energy of the model
in the thermodynamic limit. For a sufficiently large value for the
coupling constant between the $N$ two-level atoms and the single
quantized mode of the bosonic field, there is a second order phase
transition from the normal to the superradiant phase. Later,
without assuming the rotating-wave approximation and by using a
coherent state representation, the study of the stability of the
model with an infinite number of bosonic modes was presented in 
\cite{hepp2}. The study of the phase transitions in the Dicke
model was presented also by Wang and Hioe \cite{wang}, where some
of the results obtained by Hepp and Lieb were reobtained.
The generalization of the model was investigated in Refs. \cite{tese,aparicio4,tese2,tese3}, where different coupling
constants were introduced between the single mode of the bosonic field and the
ensemble of $N$ atoms, $g_{1}$ and $g_{2}$, for rotating and
counter-rotating terms, respectively. The model exhibits
a second order phase transition from the ordinary fluorescent to
the superradiant phase, respectively, at some critical temperature
$\beta^{-1}_{c}$ and also a superradiant phase
transition at zero temperature \cite{hertz} \cite{sachdev}, i.e.,
a quantum phase transition. The order parameter of the 
fluorescent to the superradiant phase transition, is the expectation value of
the excitation number associated to the bosonic mode per atom,
i.e., $lim_{N\,\rightarrow\infty}\,\frac{\langle\,b^
{\dagger}\,b\rangle}{N}\neq\,0$. In the following analysis, we will use 
functional methods to investigate the thermodynamics of the model \cite{popov2,popov}.

We would like to stress that this interpretation of the superluminescence as
a cooperative many-body phenomenon is not new in the literature. In fact, 
Mohanty and Khare \cite{mohanty} proposed that a cooperative interaction between
excited atoms and molecules is the origin of the sonoluminescence. During the compression phase of the bubble,
the temperature of the gas increases and it becomes highly excited or ionized.
The excited atoms decay as  a Dicke superradiance process. The same point of view was
proposed by Brodsky and co-workers \cite{brodsky}.
In both the above mentioned papers, the effects of the thermal bath was not taken into account.
Since it is well known that the temperature of the cavitation bubble can be as much as $25000-50000$ K,
it is fundamental to extend the study taking into account temperature effects. 
The main goal of the present work is to 
introduce finite temperature effects and further develop the interpretation of sonoluminiscence as 
a cooperative many-body effect. 

This paper is organized as follows. In
Sec. \ref{sec:gdm}, the path integral with the functional integral method is
applied to study the thermodynamics of the generalized
fermionic Dicke model.  Conclusions
are given in Sec. \ref{sec:con}. Throughout the paper we use units with $k_{B}=\hbar=1$.

\section{The generalized Dicke model and the fermionic Dicke model}
\label{sec:gdm}

\quad The Hamiltonian of a bosonic quantum system $H_{S}$, coupled with the reservoir of $N$ two-level atoms,
with Hamiltonian $H_{B}$, in thermal equilibrium at temperature
$\beta^{-1}$, can be written as
\begin{equation}
H=I_{S}\,\otimes\,H_{B}+H_{S}\,\otimes\,I_{B}+H_{I}\, , \label{13}
\end{equation}
where $I_{S}$ denote the identity in the Hilbert space of the
quantized bosonic field, $I_{B}$ denotes the identity in the
Hilbert space of the $N$ two-level atoms and $H_{I}$ is the
interaction Hamiltonian. Using the pseudo-spin operators $\sigma_{i}^{+}$,
$\sigma_{i}^{-}$ and $\sigma_{i}^z$ that satisfy the standard
angular momentum commutation relations corresponding to spin
$\frac{1}{2}$ operators, the generalized Dicke model is defined by
\begin{eqnarray}
H\,=\sum_{i=1}^{N}\,
\frac{\Omega}{2}\,\sigma_{i}^z+\omega_{0}\,b^{\dagger}\,b+\frac{g_1}{\sqrt{N}}
\sum_{i=1}^{N}\, \Bigl(b\,
\sigma_{i}^{+}+b^{\dagger}\sigma_{i}^{-}\Bigr)+\,\frac{g_2}{\sqrt{N}}
\sum_{i=1}^{N}\, \Bigl(b\,
\sigma_{i}^{-}+b^{\dagger}\sigma_{i}^{+}\Bigr)\, . \label{26}
\end{eqnarray}
In the above equation $g_1$ and $g_2$ are coupling constants
between the two-level atoms and the single mode of the bosonic field. The
$b$ and $b^{\dagger}$ are the boson annihilation and creation
operators of mode excitations that satisfy the usual commutation
relation rules.

Starting from the Hamiltonian of the Dicke model, let us define the fermionic raising
and lowering operators $\alpha^{\dagger}_{i}$, $\alpha_{i}$,
$\beta^{\dagger}_{i}$ and $\beta_{i}$, that satisfy the
anti-commutator relations
$\alpha_{i}\alpha^{\dagger}_{j}+\alpha^{\dagger}_{j}\alpha_{i}
=\delta_{ij}$ and
$\beta_{i}\beta^{\dagger}_{j}+\beta^{\dagger}_{j}\beta_{i}
=\delta_{ij}$. We can also define the following bilinear
combination of fermionic operators, $\alpha^{\dagger}_{i}\alpha_{i}
-\beta^{\dagger}_{i}\beta_{i}$, $\alpha^{\dagger}_{i}\beta_{i}$
and finally $\beta^{\dagger}_{i}\alpha_{i}$. Since the pseudo-spin
operators obey the same commutation relations as the above
presented bilinear combination of fermionic operators, we can change
the pseudo-spin operators of the Dicke model by the bilinear
combination of fermionic operators

\begin{equation}
\sigma_{i}^{z}\longrightarrow \alpha_{i}^{\dagger}\alpha_{i}
-\beta_{i}^{\dagger}\beta_{i}\, , \label{34}
\end{equation}
\begin{equation}
\sigma_{i}^{+}\longrightarrow \alpha_{i}^{\dagger}\beta_{i}\, ,
\label{35}
\end{equation}
and finally
\begin{equation}
\sigma_{i}^{-}\longrightarrow \beta_{i}^{\dagger}\alpha_{i}\, .
\label{36}
\end{equation}
With the  substitutions  defined in Eqs.
(\ref{34}), (\ref{35}) and (\ref{36}) we shall call  the
resulting Hamiltonian $H_F$ as the 
Hamiltonian of the generalized fermionic Dicke model that can be written
as
\begin{eqnarray}
H_F=\omega_0\; b^{\dagger}b+ \frac{\Omega}{2}\sum_{i=1}^N
\Bigl(\alpha_i^{\dagger}\alpha_i -\beta_i^{\dagger}\beta_i\Bigr)
+\frac{g_1}{\sqrt{N}}\sum_{i=1}^N
\Bigl(b\,\alpha_i^{\dagger}\beta_i \,+\, b^{\dagger}\,
\beta_i^{\dagger}\alpha_i \Bigr)+\,
\frac{g_2}{\sqrt{N}}\sum_{i=1}^N
\Bigl(b^{\dagger}\,\alpha_i^{\dagger}\beta_i \,+\, b\,
\beta_i^{\dagger}\alpha_i \Bigr)\, . \label{37}
\end{eqnarray}

Next, we consider the problem of defining the
partition function $Z_F$ of the generalized fermionic Dicke model.
First let us define the Euclidean action $S$ of this model, which
describes a single quantized mode of the field and the ensemble of
$N$ identical two-level atoms. The Euclidean action $S$ is given by
\begin{equation}
S=\int_0^{\beta} d\tau \biggl(b^*(\tau)\frac{\partial}{\partial
\tau} b(\tau)+ \sum_{i=1}^{N}
\Bigl(\alpha^*_i(\tau)\frac{\partial}{\partial \tau}\alpha_i(\tau)
+\beta^*_i (\tau)\frac{\partial}{\partial
\tau}\beta_i(\tau)\Bigr)\biggr) -\int_0^{\beta}d\tau H_{F}(\tau)\,
\label{66}
\end{equation}
where $H_{F}$ is the full Hamiltonian for the generalized fermionic
Dicke model given by
\begin{eqnarray}
H_{F}\,=\,\omega_{0}\,b^{\,*}(\tau)\,b(\tau)\,+
\,\frac{\Omega}{2}\,\displaystyle\sum_{i\,=\,1}^{N}\,
\biggl(\alpha^{\,*}_{\,i}(\tau)\,\alpha_{\,i}(\tau)\,-
\,\beta^{\,*}_{\,i}(\tau)\beta_{\,i}(\tau)\biggr)\,+
\nonumber\\
+\,\frac{g_{\,1}}{\sqrt{N}}\,\displaystyle\sum_{i\,=\,1}^{N}\,
\biggl(\alpha^{\,*}_{\,i}(\tau)\,\beta_{\,i}(\tau)\,b(\tau)\,+
\alpha_{\,i}(\tau)\,\beta^{\,*}_{\,i}(\tau)\,b^{\,*}(\tau)\,\biggr)\,+
\nonumber\\
+\,\frac{g_{\,2}}{\sqrt{N}}\,\displaystyle\sum_{i\,=\,1}^{N}\,
\biggl(\alpha_{\,i}(\tau)\,\beta^{\,*}_{\,i}(\tau)\,b(\tau)\,+
\,\alpha^{\,*}_{\,i}(\tau)\,\beta_{\,i}(\tau)\,b^{\,*}(\tau)\biggr).
\label{66a}
\end{eqnarray}

To proceed, let us define the formal quotient of two functional integrals,
i.e., the partition function of the generalized fermionic Dicke
model and the partition function of the free fermionic Dicke model.
Therefore we are interested in calculating the following quantity
\begin{equation}
\frac{Z_{F}}{Z_{F_{0}}}=\frac{\int [d\eta]\,e^{\,S}}{\int
[d\eta]\,e^{\,S_{0}}}\, , \label{65}
\end{equation}
where $S$
is the Euclidean action of the generalized fermionic Dicke model
given by Eq. (\ref{66}),
$S_0$
is the free Euclidean action for the free single bosonic mode and
the free two-level atoms.
The functional integrals involved in Eq. (\ref{65}), are
functional integrals with respect to the complex functions
$b^*(\tau)$ and $b(\tau)$ and Grassmann fermionic fields
$\alpha_i^*(\tau)$, $\alpha_i(\tau)$, $\beta_i^*(\tau)$ and
$\beta_i(\tau)$. Since we are using thermal equilibrium boundary
conditions, in the imaginary time formalism \cite{matsubara,kubo,martin}, the integration
variables in Eq. (\ref{65}) obey periodic boundary conditions for
the Bose field, i.e., $b(\beta)=b(0)$ and anti-periodic boundary
conditions for fermionic fields i.e., $\alpha_i(\beta)=-\alpha_i(0)$
and $ \beta_i(\beta)=-\beta_i(0)$.

The free action for the single mode bosonic field $S_{0}(b)$ is
given by
\begin{equation}
S_{0}(b) = \int_{0}^{\beta} d\tau \biggl(b^{*}(\tau)
\frac{\partial b(\tau)}{\partial \tau} -
\omega_{0}\,b^{*}(\tau)b(\tau)\biggr)\, . \label{67}
\end{equation}
Then we can write the action $S$ of the generalized fermionic Dicke
model, given by Eq. (\ref{66}), using the free action for the
single mode bosonic field $S_{0}(b)$, given by Eq. (\ref{67}), plus
an additional term that can be expressed in a matrix form.
Therefore the total action $S$ can be written as
\begin{equation}
S = S_{0}(b) +  \int_{0}^{\beta} d\tau\,\sum_{i=1}^{N}\,
\rho^{\dagger}_{i}(\tau)\, M(b^{*},b)\,\rho_{i}(\tau)\, ,
\label{68}
\end{equation}
where $\rho_{\,i}(\tau)$ is a column matrix given in terms of
fermionic field operators
\begin{eqnarray}
\rho_{\,i}(\tau) &=& \left(
\begin{array}{c}
\beta_{\,i}(\tau) \\
\alpha_{\,i}(\tau)
\end{array}
\right),
\nonumber\\
\rho^{\dagger}_{\,i}(\tau) &=& \left(
\begin{array}{cc}
\beta^{*}_{\,i}(\tau) & \alpha^{*}_{\,i}(\tau)
\end{array}
\right) \label{69a}
\end{eqnarray}
and the matrix $M(b^{*},b)$ is given by
\begin{equation}
M(b^{*},b) = \left( \begin{array}{cc}
L & -(N)^{-1/2}\,\biggl(g_{1}\,b^{*}\,(\tau) + g_{2}\,b\,(\tau)\biggr)\\
-(N)^{-1/2}\,\biggl(g_{1}\,b\,(\tau) + g_{2}\,b^{*}\,(\tau)\biggr)
&
L_*
\end{array} \right)\,.
\label{69b}
\end{equation}
The operators $L$ and $L_*$ are defined by $\partial_{\tau} + \Omega/2$ and
$\partial_{\tau} - \Omega/2$, respectively.
Substituting the action $S$ given by Eq. (\ref{68}) in the functional integral form of the partition
function, we see that this functional integral is Gaussian in the fermionic fields.
Now, let us begin integrating with respect to these fermionic fields; therefore we obtain
\begin{eqnarray}
Z=\int[d\eta(b)]\,e^{S_{B0}} \Bigl(\det{M(b^{*},b)}\Bigr)^N\,,
\label{Zop}
\end{eqnarray}
where, in this case, $[d\eta(b)]$ is the functional measure only for the bosonic field.
With the help of the following property for matrices with operator components
\begin{eqnarray}
\det\left(\begin{array}{cc}
A&B\\
C&D
\end{array}
\right)=\det\left(AD-ACA^{-1}B\right)\,,
\label{mazprop}
\end{eqnarray}
and the determinant properties, we have that
\begin{eqnarray}
\det{M(b^{*},b)}=\det{\Bigl(LL_*\Bigr)}\,\det{\left(1-N^{-1}L_*^{-1}
\Bigl(g_1\,b+ g_2\,b^*\Bigr)L^{-1}\Bigl(g_1\,b^*+ g_2\,b\Bigr)\right)}\,.
\label{Mop1}
\end{eqnarray}
Substituting Eq. (\ref{Zop}) and Eq. (\ref{Mop1}) in Eq. (\ref{65}), we have that
\begin{eqnarray}
\frac{Z}{Z_0}=\frac{Z_A}{\int[d\eta(b)]\,e^{S_{B0}}}\,,
\label{ZA0}
\end{eqnarray}
with $Z_A$ defined by
\begin{eqnarray}
Z_A=\int[d\eta(b)]\exp{\left(S_{B0}+N\,tr\ln\biggl(1-N^{-1}L_*^{-1}
\Bigl(g_1\,b+ g_2\,b^*\Bigr)L^{-1}\Bigl(g_1\,b^*+ g_2\,b\Bigr)\biggr)\right)}\,.
\label{ZA1}
\end{eqnarray}

We are interested in knowing the asymptotic behaviour of the quotient $\frac{Z}{Z_0}$ in the
thermodynamic limit, i. e., $N\rightarrow\infty$. With this intention, we analyse
the asymptotic behaviour of the last defined expression $Z_A$. First, let us scale the bosonic
field by $b\rightarrow\sqrt{N}\,b$ and $b^*\rightarrow\sqrt{N}\,b^*$, so that we get
\begin{eqnarray}
Z_A=A(N)\int[d\eta(b)]\exp{\left(N\,\Phi(b^*,b)\right)}\,,
\label{ZA2}
\end{eqnarray}
with the function $\Phi(b^*,b)$ defined by
\begin{eqnarray}
\Phi(b^*,b)=S_{B0}+tr\ln\biggl(1-L_*^{-1}
\Bigl(g_1\,b+ g_2\,b^*\Bigr)L^{-1}\Bigl(g_1\,b^*+ g_2\,b\Bigr)\biggr)\,.
\label{fi1}
\end{eqnarray}
The term $A(N)$ in Eq. (\ref{ZA2}) comes from transforming the functional measure $[d\eta(b)]$ under scaling
the bosonic field by $b\rightarrow\sqrt{N}\,b$ and $b^*\rightarrow\sqrt{N}\,b^*$. The asymptotic behaviour
of the integral functional appearing in Eq. (\ref{ZA2})
when $N\rightarrow\infty$, can be obtained by using the method of steepest descent. In this method,
we expand the function $\Phi(b^*,b)$ around the point $b(\tau)=b_0(\tau)$ and
$b^*(\tau)=b^*_0(\tau)$, which can be of two kinds. One kind that makes $Re(\Phi(b^*,b))$ maximum,
and the other kind is defined as the saddle point.
We consider the first terms of the expansion in the integral functional, which are the leading
terms for the value of the integral function. We can find the maximum points, or
saddle points, finding the stationary points. The stationary points are solution of the following equations $\frac{\delta\,
\Phi(b^*,b)}{\delta\,b(\tau)}=0$ and $\frac{\delta\,\Phi(b^*,b)}{\delta\,b^*(\tau)}=0$.
For the full Dicke model, the stationary points are constant functions $b(\tau)=b_0$ and $b^*(\tau)=b^*_0$.
It is not difficult to show that for $\beta\leq\beta_c$ the stationary point
is given by $b_0=b_0^*=0$, which is a maximum point. The critical value $\beta_c$ is obtained by solving the following equation
\begin{eqnarray}
\frac{\omega_0\,\Omega}{(g_1+g_2)^2}=\tanh\left(\frac{\beta_c\,\Omega}{2}
\right)\,.
\label{tcrit}
\end{eqnarray}
In this last equation, it is possible to find some solution for $\beta_c$, in the
case of $(g_1+g_2)^2>\omega_0\Omega$, which corresponds to a strong-coupling regime. With this condition the system undergoes a phase transition.
When the system has $\beta<\beta_c$ we say that the system is in the normal phase.
For $\beta>\beta_c$ the stationary points $b(\tau)=b_0$ and $b^*(\tau)=b^*_0$ satisfy the following
equation
\begin{eqnarray}
\frac{\omega_0\,\Omega_{\Delta}}{(g_1+g_2)^2}=\tanh\left(\frac{\beta\,\Omega_{\Delta}}{2}\right)\,,
\label{tra1}
\end{eqnarray}
with $\Omega_{\Delta}$ defined by
\begin{eqnarray}
\Omega_{\Delta}=
\sqrt{\Omega^2+4\,(g_1+g_2)^2\,|b_0|^2}\,.
\label{omegadelta}
\end{eqnarray}
Phase transition happens if it is possible to find some real solution for $|b_0|\neq 0$ in Eq. (\ref{tra1}).
It is only possible when $(g_1+g_2)^2>\omega_0\,\Omega$ and $\beta>\beta_c$. In the case of $g_1\neq 0$ and
$g_2=0$, and also in the case of $g_1=0$ and $g_2\neq 0$, the maximum points are a continuous set of values given
by the expression $b_0=\rho\,e^{i\,\phi}$ and $b^*_0=\rho\,e^{-i\,\phi}$ with $\phi\in [0,2\pi)$ and
$\rho=|b_0|$, with $|b_0|$ defined by Eq. (\ref{tra1}). In the case of $g_1\neq 0$ and $g_2\neq 0$, we have
two maximum points, which are given by $b^*_0=b_0=\pm |b_0|$, with $|b_0|$ defined by
Eq. (\ref{tra1}). When the system has $\beta>\beta_c$ we say that the system is in the superradiant phase.

After these calculations, we can discuss if there are experimentally accessible systems where superluminescence could be
explained as a cooperative many-body phenomenon. It is well known that the temperature of the cavitation bubble can be as much as $25000-50000$ K.
Taking into account temperature effects, from the above discussion we see that for excited systems that couple strongly
to an electromagnetic field it is possible to have cooperative radiative processes. We must have
\begin{equation}
{(\,g_{\,1}+
g_{\,2}\,)^{\,2}}\gg\omega_{0}\,\Omega.
\label{ultrastrong}
\end{equation}
In conclusion, superluminescence as
a cooperative many-body phenomenon may happen in a ultrastrong-coupling regime between the
two-level atoms and the electromagnetic field.


It is known that at zero temperature the generalized Dicke model (Eq. (\ref{26})) exhibits a collective spontaneous emission (superradiance phenomena). This superradiation happens when the values of the coupling constants between atoms and the bosonic mode satisfy the inequality $(g_1+g_2)^2 > \omega_0 \Omega$  (the strong-coupling regime). An important feature characterizing the superradiance phenomena is that this happens with transitions between collective states of the atoms, called Dicke states. Another way to the Dicke model be superradiant is allowing the system (atoms and bosonic mode) be in contact with a thermal bath. It was shown (see Refs. \cite{hepp,hepp2,wang}) that the model exhibits a phase transition at finite temperature, where according to Eq. (\ref{tcrit}) a superradiant phase appears when $(g_1+g_2)^2 > \omega_0 \Omega$, and the critical temperature is obtained from the same Eq. (\ref{tcrit}). Above the critical temperature the system is in the normal phase, where the atomic polarization $\left\langle J_z\right\rangle$ and  $\left\langle b^{\dagger}b\right\rangle$ equal to zero. In the opposite case, below the critical temperature, the system is in the superradiant phase, where $\left\langle J_z\right\rangle$ and  $\left\langle b^{\dagger}b\right\rangle$ are different from zero. It is important to emphasize that, at finite temperature, transitions between atoms states not only happen between collective Dicke states but rather between another general separable states. Even in such case of incomplete coherence between the atoms, there appear the superradiance phenomena.

The physical system where sonoluminescene appears is in the context of a Dicke model in contact with a thermal bath. Therefore in Eq. (\ref{tcrit}) we see that the condition to obtain a superradiant phase is 
$(g_1+g_2)^2 > \omega_0 \Omega$. Experimentally such condition of strong-coupling between atoms and the bosonic field is difficult to achieve. Increasing the temperature at the level $\beta\ll 1$ ($K_B T\gg 1$) we have $\tanh(\beta \Omega/2)\ll 1$ and from Eq. (\ref{tcrit}) we obtain that $(g_1+g_2)^2 \gg \omega_0 \Omega$, where such condition is even more difficult to achieve. This result shows that it is more difficult to obtain a superradiant phase when the temperature increase, whatever in future studies the superradiant condition  $(g_1+g_2)^2 \gg \omega_0 \Omega$ could be explored inside the bubble.

In the literature we can find works studying the sonoluminescence effect where light emission is obtained as a lasing process in atoms where collective transitions between them are present (see Refs. \cite{moha1,moha2}). The author considers interaction between the electromagnetic field and Argon eximers present in the bubbles. In contrast with our thermal equilibrium study of the model, in order to study lasing processes, this author studied temporal evolution equations called  Maxwell-Bloch equations, which include finite temperature effects.


\section{Conclusions}
\label{sec:con}

Sonoluminescence is a process in which a strong sound field is used in a 
liquid to produce light. In this work, we presented a theoretical description 
of this phenomenon based upon the Dicke superradiance.  We used the interpretation
of sonoluminescence as a cooperative many-body phenomenon by studying a generalized 
fermionic Dicke model in which finite temperature effects are taking into account. This is 
an essential issue since it has been experimentally proved that the temperature of the 
cavitation bubble is finite and can be as much as $25000-50000$ K.  At a certain critical temperature,
the model is characterized by a second order phase transition during which the corresponding system passes
from the ordinary fluorescent phase to a superradiant phase, explaining the light production.  
From the analysis of the expression for the critical temperature we see that superluminescence 
can take place in the strong-coupling regime between the atoms and the electromagnetic field.
The main conclusion of this work is that sonoluminescence can be interpreted as a 
finite-temperature many-body effect

There
are some possible generalizations of the results presented here. First,  instead of two-level atoms we can consider
$N$-level atoms. The problem of superradiance for many-level systems was investigated by
Shelepin \cite{shelepin}.  In the case of three, four and more levels, the states of the
system must be described by the groups $SU(3)$, $SU(4)$, and so on. The cooperative behavior
among three-level systems in the presence of a bosonic field was investigated by
Bowden and Sung \cite{bowden}. More recently, the superradiance of three-level systems
was investigated in \cite{emary1, emary2}.
We would like to stress that we studied a simplified model with only one mode of the
field coupled to two-level atoms. Some important effects that cannot be studied in the single-mode
model is the angular distribution of the radiation and also the role played by the
geometric configuration of the medium in the phenomenon. 
For a better description of this phenomenon, a multi-mode theory and the
generalization for $N$-level atoms is under investigation.

Another interesting aspect of the Dicke model is in the context of information geometry. Recently, it was shown that the equilibrium space of the Dicke model does not contain information about the quantum phase transitions of the system \cite{tapo}, indicating that information geometry cannot be applied. In other similar cases, this problem has been solved by using the formalism of geometrothermodynamics \cite{quev07}; it would interesting to find out if this is also possible in the case of the quantum and finite-temperature phase transitions of the generalized Dicke model. If so, sonoluminescence could also be investigated geometrically in the context of geometrothermodynamics.

\section{Acknowledgements}

We would like to thank E. Arias and C. H. G. Bessa for useful discussions. 
MAA thanks the Instituto de F\'{\i}sica Te\'orica (IFT-UNESP) for their kind
hospitality. 
HQ thanks the members of the ICRA-group for stimulating discussions and excellent hospitality during his stay at CBPF.  
MAA was partially supported by FAPESP and NFS and HQ were supported by CNPq.


\begin{thebibliography}{99}

\bibitem{sol1} H. Frenzel and H. Schultes, Z. Phys. Chem. {\bf B27}, 421 (1934).
\bibitem{sol2} B. P. Barber, R. A. Hiller, R. L\"ofstedt, S. J. Putterman and K. R. Weninger,
Phys. Rep. {\bf 281}, 65 (1997).
\bibitem{margulis} M. A. Margulis, Physics-Uspekhi {\bf 43}, 259 (2000).
\bibitem{sol3} D. F. Gatan, L. A. Crum, C. C. Church and R. A. Roy, J. Acoust. Soc. Am. {\bf 91}, 3166 (1992).
\bibitem{sol4} M. P. Brenner, S. Hilgenfeldt and D. Lohse, Re. Mod. Phys. {\bf 74}, 425 (2002).
\bibitem{sol5} L. Rayleigh, Phyl. Mag. {\bf 34}, 94 (1917).
\bibitem{sol6} M. S. Plesset, Jour. App. Phys. {\bf 25}, 96 (1954).
\bibitem{sol7} M. S. Plesset and A. Prosperetti, Ann. Rev. Fluid Mech. {\bf 9}, 145 (1977).
\bibitem{sol8} B. P. Barber and S. J. Putterman, Phys. Rev. Lett. {\bf 69}, 3839 (1992).
\bibitem{sol9} C. C. Wu and P. H. Roberts, Proc. R. Soc. Lond. Ser. {\bf A445}, 323 (1994).
\bibitem{sol10} C. Eberlein, Phys. Rev. {\bf A53}, 2772 (1996).
\bibitem{sol11} C. Eberlein, Phys. Rev. Lett. {\bf 76}, 3842 (1996).
\bibitem{sol12} C. S. Unnikrishnan and S. Mukhopadhyay, Phys. Rev. Lett. {\bf 77}, 4690 (1996).
\bibitem{sol13} A. Lambrecht, M. T. Jaekel and S. Reynaud, Phys. Rev. Lett. {\bf 78}, 2267 (1997).
\bibitem{sol14} N. Garcia and A. P. Levanyuk, Phys. Rev. Lett. {\bf 78}, 2268 (1997).
\bibitem{sol15} R. Hiller, S. J. Putterman and B. P. Barber, Phys. Rev. Lett. {\bf 69}, 1182 (1992).
\bibitem{sol16} R. L\"ofstedt, B. P. Barber and S. J. Putterman, Phys. Fluids {\bf A 5} 2911 (1993).
\bibitem{sol17} L. Frommhold, Phys. Rev. {\bf E58}, 1899 (1998).
\bibitem{sol18} R. H. Dicke, Phys. Rev. {\bf 93}, 99 (1954).
\bibitem{hioe} F. T. Hioe, Phys. Rev. {\bf A8}, 1440 (1973).
\bibitem{carmichael} H. J. Carmichael, C. W. Gardiner and D. F.
Walls, Phys. Lett. {\bf A46}, 47 (1973).
\bibitem{duncan} G. Comer Duncan, Rev. {\bf A9}, 418 (1974).
\bibitem{sol19} A. V. Andreev, V. I. Emel'yanov and Y. A. II'inskii, Sov. Phys. Usp. {\bf 23}, 493 (1980).
\bibitem{yarunin} V. B. Kir'yanov and V. S. Yarunin, Theor. Math.
Phys. {\bf 43}, 340 (1980).
\bibitem{sol20} M. Gross and S. Haroche, Phys. Rep. {\bf 93}, 301 (1982).
\bibitem{garraway} B. M. Garraway, Phil. Trans. R. Soc. {\bf A369}, 1137 (2011).
\bibitem{hepp} K. Hepp and E. H. Lieb, Ann. Phys. {\bf 76}, 360 (1973).
\bibitem{hepp2} K. Hepp and E. H. Lieb, Phys. Rev. {\bf A8}, 2517 (1973).
\bibitem{wang} Y. K. Wang and F. T. Hioe, Phys. Rev. {\bf A7}, 831 (1973).
\bibitem{tese} M. Aparicio Alcalde, A. L. L. de Lemos and N. F. Svaiter, J. Phys. {\bf A40}, 11961 (2007).
\bibitem{aparicio4} M. Aparicio Alcalde and B. M. Pimentel, Physica A {\bf 390}, 3385 (2011).
\bibitem{tese2} M. Aparicio Alcalde, R. Kullock and N. F. Svaiter, J. Math. Phys. {\bf 50}, 013511-1 (2009).
\bibitem{tese3} M. Aparicio Alcalde, S. Stephany and N. F. Svaiter, J. Phys. {\bf A44}, 505301 (2011).
\bibitem{hertz} J. A. Hertz, Phys. Rev. {\bf B14}, 1165 (1976).
\bibitem{sachdev} S. Sachdev, {\em{"Quantum Phase Transitions"}}, Cambridge University Press,
Cambridge (1999).
\bibitem{popov2} V. N. Popov and S. A. Fedotov, Theor. Math.
Phys. {\bf 51}, 363 (1982).
\bibitem{popov} V. N. Popov and S. A. Fedotov, Sov. Phys. Jetp. {\bf 67},
535 (1988).
\bibitem{mohanty} P. Mohanty and S. V. Khare, Phys. Rev. Lett. {\bf 80}, 189 (1998).
\bibitem{brodsky} A. M. Brodsky, L. W. Burgess and A. L. Robinson, Ultrasonic {\bf 39} 97 (2001).
\bibitem{matsubara} T. Matsubara, Prog. Theor. Phys. {\bf 14}, 351
(1955).
\bibitem{kubo} R. Kubo, J. Phys. Soc. Jap. {\bf 12}, 570 (1957).
\bibitem{martin} P. C. Martin and J. Schwinger, Phys. Rev. {\bf 115}, 1342 (1959).
\bibitem{shelepin} L. A. Shelepin, Sov. Phys. JETP {\bf 27}, 784 (1968).
\bibitem{bowden} C. M. Bowden and C. C. Sung, Phys. Rev. {\bf A18}, 1558 (1978).
\bibitem{emary1} M. Hayn, C. Emary and T. Brandes, Phys. Rev. {\bf A84}, 053856 (2011).
\bibitem{emary2}  M. Hayn, C. Emary and T. Brandes, Phys. Rev. {\bf A86}, 063822 (2012).
\bibitem{tapo} A. Dey, S. Mahapatra, P. Roy, and T. Sarkar, Phys. Rev. {\bf E86}, 031137 (2012).
\bibitem{quev07} H. Quevedo, J. Math. Phys. {\bf 48}, 013506 (2007).
\bibitem{moha1} P. Mohanty, {\it An Analytic Description of Light Emission in Sonoluminescence}, arXiv: cond-mat/9912271.
\bibitem{moha2} P. Mohanty, {\it Electromagnetic Field Correlation inside a Sonoluminescing Bubble}, arXiv: cond-mat/0005233.

\end{thebibliography}
\end{document}